# TATA KELOLA DATABASE PERGURUAN TINGGI YANG OPTIMAL DENGAN DATA WAREHOUSE

**Spits Warnars**
Fakultas Teknologi Informasi, Universitas Budi Luhur
Jl. Petukangan Selatan, Kebayoran Lama, Jakarta 12260, 021-5853753
e-mail: spits@bl.ac.id

*Abstract*

*The emergence of new higher education institutions has created the competition in higher education market, and data warehouse can be used as an effective technology tools for increasing competitiveness in the higher education market. Data warehouse produce reliable reports for the institution's high-level management in short time for faster and better decision making, not only on increasing the admission number of students, but also on the possibility to find extraordinary, unconventional funds for the institution. Efficiency comparison was based on length and amount of processed records, total processed byte, amount of processed tables, time to run query and produced record on OLTP database and data warehouse. Efficiency percentages was measured by the formula for percentage increasing and the average efficiency percentage of 461.801,04% shows that using data warehouse is more powerful and efficient rather than using OLTP database. Data warehouse was modeled based on hypercube which is created by limited high demand reports which usually used by high level management. In every table of fact and dimension fields will be inserted which represent the loading constructive merge where the ETL (Extraction, Transformation and Loading) process is run based on the old and new files.*

*Abstrak*

*Pertumbuhan perguruan tinggi menimbulkan persaingan pada pasar perguruan tinggi dan data warehouse dapat digunakan sebagai sebuah senjata teknologi untuk bersaing dalam pasar perguruan tinggi. Data warehouse menghasilkan dalam waktu yang singkat laporan yang dapat dipercaya bagi manajemen tingkat atas perguruan tinggi didalam membuat keputusan yang cepat dan terbaik dan tidak hanya menambah jumlah mahasiswa akan tetapi kemungkinan untuk mendapatkan dana atau investasi yang tak pernah terpikirkan berdasarkan aturan perguruan tinggi. Perbandingan prosentasi efisiensi akan diukur berdasarkan pada total byte yang dikelola, record yang dikelola, panjang record yang diproses, jumlah tabel yang diproses, waktu dan record yang dihasilkan pada database OLTP dan data warehouse. Prosentase efisiensi diukur dengan rumus kenaikan prosentase dan rata-rata efisiensi kenaikan prosentase 461.801,84%, menunjukkan bahwa penggunaan data warehouse lebih handal dan efisien dibandingkan penggunaan database OLTP. Data warehouse dimodelkan dengan hypercube yang terbentuk dan dibatasi berdasarkan laporan-laporan yang sering dipakai oleh manajemen tingkat atas dan didalam setiap tabel fakta dan dimensi akan diberikan field yang mencirikan metode Loading constructive merge dimana proses ETL (Extraction, Transformation and Loading) akan dijalankan berdasarkan perbedaan berkas lama dan baru.*

***Kata Kunci:*** *Data warehouse, Perguruan tinggi, Hypercube, konsep dimensi bisnis*

## 1. PENDAHULUAN

Persaingan untuk mendapatkan mahasiswa antar perguruan tinggi tidak dapat dipelakkan lagi seiring dengan pertumbuhan perguruan tinggi [1] yang semakin pesat dan setiap peguruan tinggi harus memperlengkapi dirinya untuk dapat bertahan didalam persaingan yang pada akhirnya perguruan tinggi yang tidak peka akan kalah bersaing dan menutup usahanya. Di tangan yang handal dan tepat database sebagai teknologi penyimpanan transaksi harian yang permanent akan dapat dirubah menjadi sebuah senjata teknologi yang handal untuk menang didalam menghadapi persaingan [2-3]. Tidak pada tempatnya lagi perguruan tinggi hanya mengejar keuntungan semata-mata dengan mengabaikan teknologi yang seharusnya bisa membantu proses bisnis didalam perguruan tinggi tersebut. Dimana biasanya kekalahan didalam persaingan perguruan tinggi hanya bisa menyalahkan manajemen tingkat atas yang tidak pernah diperlengkapi dengan teknologi dan hanya mengandalkan hal yang bersifat tebak-tebak atau perkiraan saja. Tuntutan masyarakat untuk transparansi pendidikan perguruan tinggi dapat terwujud dan memudahkan Dirjen pendidikan tinggi memantau perguruan tinggi secara transparansi [4].

Data yang ada pada perguruan tinggi yang begitu besar dan banyak membutuhkan alokasi tempat penyimpanan akan terbantu dan menjadi efisien dengan adanya data warehouse [5]. Dengan adanya data warehouse menangkap seluruh data bisnis proses yang ada dari mulai yang berhadapan dengan mahasiswa sebagai konsumen, proses pengajaran dan keseluruhan sistem informasi yang ada dalam perguruan tinggi [6].

Tak pelak lagi sudah saatnya data warehouse harus diimplementasikan pada perguruan tinggi dan sudah banyak yang menerapkannya [7], bahkan data warehouse digunakan pada proses pembelajaran sebagai proses utama didalam perguruan tinggi [8-10] bahkan lebih dari itu mengarah ke *Data mining* [11]. Berbeda dengan penelitian pada [8-10] dimana hanya terbatas pada pentingnya penerapan data warehouse pada sebuah universitas, maka penulisan ini akan menekankan dengan nilai kuantitatif sebagai pembuktian bahwa penggunaan data warehouse lebih handal dan efektif dibandingkan database OLTP. Selain itu efisiensi pada bentukan *hypercubes* akan dijelaskan pada masing-masing pembentukan *hypercubes* dengan mengurangi tabel dimensi dan menyatukan dalam sebuah tabel dimensi sebagai sebuah improvisasi untuk peningkatan kinerja *query* sewaktu mengakses data warehouse.

## 2. METODE PENELITIAN

Penelitian ini adalah penelitian deksriptif kuantitatif, dimana hasil penelitian akan menjelaskan prosedur pengimplementasian data warehouse, tanpa melakukan perbandingan ataupun dihubungkan dengan penelitian lainnya dan hasil prosentase kuantitatif dari penelitian ini sebagai penegas keberhasilan penelitian ini dalam bentuk angka, bahwa penggunaan data warehouse lebih handal dibandingkan menggunakan database OLTP.

Penelitian ini dilakukan dengan langkah-langkah berikut:
1) Pembatasan masalah dengan hanya mengumpulkan 5 laporan yang sering digunakan oleh manajemen tingkat atas
2) Mendesain *hypercubes* dari masing-masing laporan
3) Mendesain model data logika data warehouse berdasarkan hypercubes yang terbentuk.
4) Implementasi database OLTP dengan menggunakan aplikasi database MySQL berdasarkan pembatasan pada pembuatan kelima laporan di atas
5) Implementasi database data warehouse dengan menggunakan aplikasi database MySQL.
6) Proses eksekusi query dengan menggunakan database OLTP untuk menghasilkan kelima laporan di atas.
7) Proses eksekusi query dengan menggunakan database data warehouse untuk menghasilkan kelima laporan di atas.
8) Pembandingan efisiensi proses antara database OLTP dan data warehouse berdasarkan parameter sebagai berikut:
    a. Total byte yang dikelola.
    b. Record yang dikelola.
    c. Panjang record yang diproses.
    d. Jumlah tabel yang diproses.
    e. Waktu menjalankan query.
    f. Record yang dihasilkan dari query.
9) Nilai kuantitatif dibuktikan dengan efisiensi dengan tren prosentase kenaikan dengan rumus:
    (data lama–data baru)/data baru * 100

Nilai kuantitatif pada langkah kesembilan metode penelitian ini akan membuktikan bahwa penggunaan data warehouse lebih efektif daripada database pada umumnya. Penelitian lain menggunakan matrik untuk menguji kualitas multidimensional model dimana matrik tersebut diujikan pada tabel star dan skema pada data warehouse berdasarkan jumlah field, jumlah foreign key pada tabel-tabel tersebut [12].

Pada tulisan ini pendekatan yang akan dipakai dalam pengembangan model dimensi adalah mengacu pada pengguna dengan pertimbangan sebagai berikut:
1) Memuaskan kebutuhan berbagai tingkat manajemen, terutama manajemen tingkat menengah sampai manajemen tingkat atas.
2) Mendapatkan kebutuhan laporan dan mendesain d*ata warehouse* yang memang dibutuhkan oleh manajemen.

**2.1. Pengumpulan 5 laporan yang sering digunakan oleh manajemen tingkat atas**
Langkah pertama sesuai dengan penjelasan sebelumnya sebelumnya dimana untuk lebih memudahkan dalam tingkat kerumitan data dan laporan-laporan yang dikelola, sebagai contoh akan digunakan 5 buah laporan-laporan yang sering digunakan atau dibutuhkan oleh manajemen tingkat atas perguruan tinggi. Adapun laporan-laporan tersebut adalah: (1) laporan jumlah mahasiswa per jenjang program studi per jenis kelamin per angkatan, (2) laporan jumlah mahasiswa aktif per semester tahun ajaran per jenjang program studi per jenis kelamin per angkatan, (3) laporan jumlah komposisi indeks prestasi per semester tahun ajaran per jenjang program studi per angkatan, (4) laporan jumlah komposisi grade nilai per semester tahun ajaran per jenjang program studi per jenis kelamin per angkatan, dan (5) laporan pengajaran dosen per semester tahun ajaran. Untuk menghasilkan kelima laporan di atas digunakan database OLTP yang terdapat pada Gambar 1.

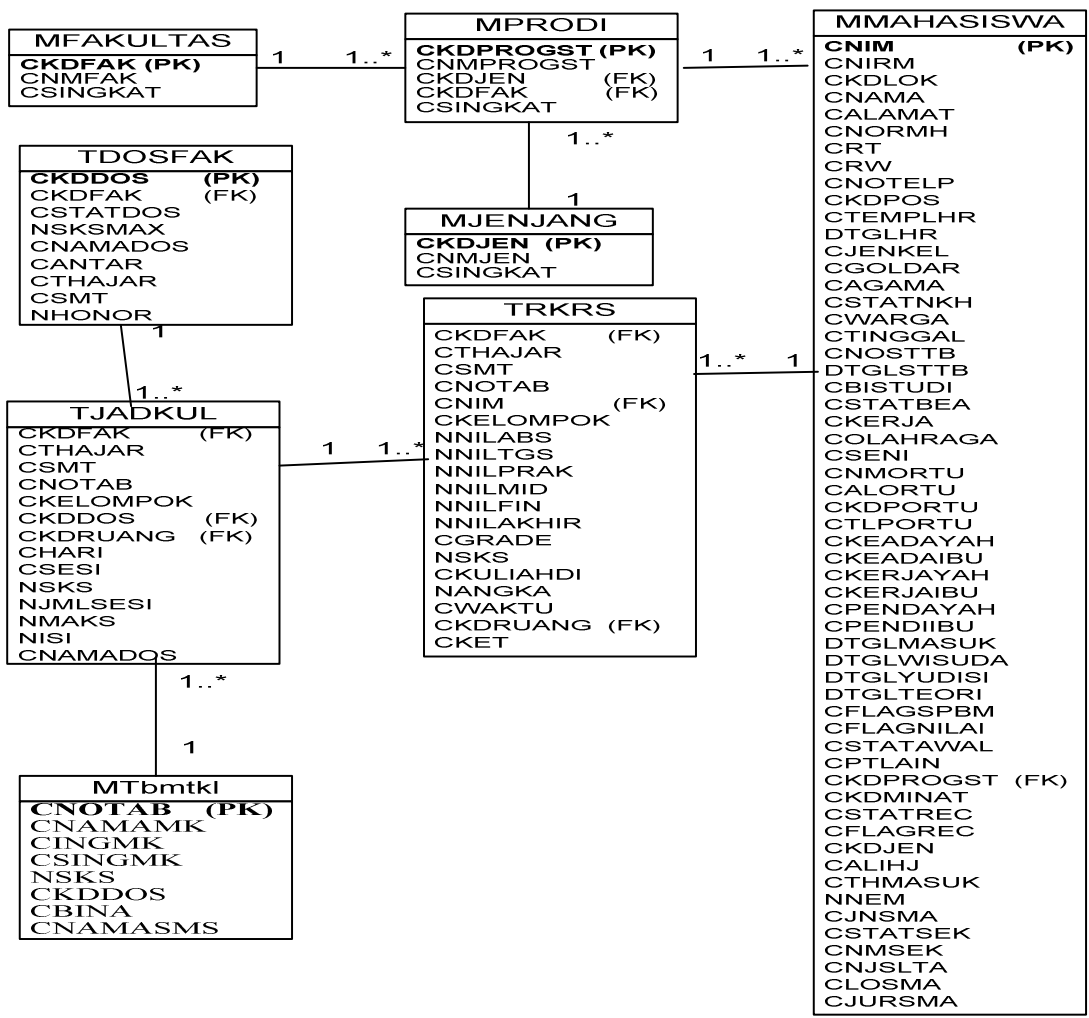

Gambar 1  *Class diagram* model data logika *OLTP*

**2.2. Desain Hypercubes**

Langkah kedua adalah mendesain *hypercubes* masing-masing kelima pembatasan laporan di atas dan efisiensi akan dilakukan pada masing-masing hypercubes, dimana tabel dimensi akan dikurangi dan disatukan pada sebuah tabel dimensi dengan tujuan untuk peningkatan kinerja. Eliminasi dimensi akan dilakukan pada dimensi jenis kelamin, angkatan, jenjang studi, tahun ajaran, semester, *grade*, mata kuliah, dosen dan kelompok indeks prestasi. Eliminasi dimensi akan dijelaskan pada masing-masing pembentukan *hypercube*. Gambar 2 memperlihatkan *hypercube* laporan jumlah mahasiswa per jenjang program studi per jenis kelamin per angkatan. Berdasarkan *hypercube* Gambar 2 maka akan terbentuk 1 tabel fakta dan 4 tabel dimensi, namun untuk dimensi Jenis kelamin karna hanya berisi dengan 2 nilai yaitu Pria atau Wanita,maka kita tidak perlu membuat tabel dimensi untuk dimensi jenis kelamin ini. Hal ini juga sama dengan dimensi angkatan, dimana dimensi angkatan ini hanya berisi tahun masuk mahasiswa. Demikian juga dengan dimensi jenjang studi dimana dimensi jenjang studi ini hanya berisi kode 50 untuk jenjang studi strata satu dan kode 30 untuk jenjang studi diploma 3. Oleh karena itu akan terbentuk 1 tabel fakta dan 1 tabel dimensi, hal ini akan terlihat pada model data Gambar 3. Tabel fakta adalah WData1 dan tabel dimensi adalah WPRODI. Tabel dimensi WPRODI adalah penggabungan 3 tabel pada sistem *OLTP* yaitu tabel MPRODI, MFAKULTAS dan MJENJANG. Akhirnya jika menggunakan d*ata warehouse* maka laporan ini membutuhkan 2 tabel basis data yaitu WData1 dan WPRODI dimana jika menggunakan database OLTP menggunakan 4 tabel yaitu MMAHASISWA, MPRODI, MFAKULTAS dan MJENJANG.

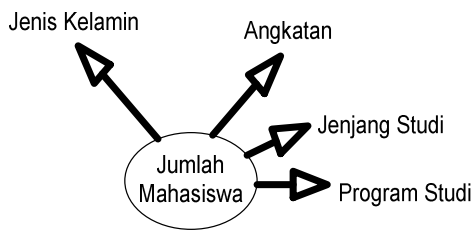

Gambar 2. *Hypercube* laporan jumlah mahasiswa per jenjang program studi per jenis kelamin per angkatan

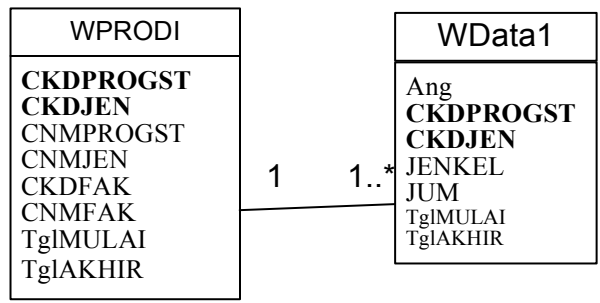

Gambar 3. *Class diagram data warehouse* laporan jumlah mahasiswa per jenjang program studi per jenis kelamin per angkatan

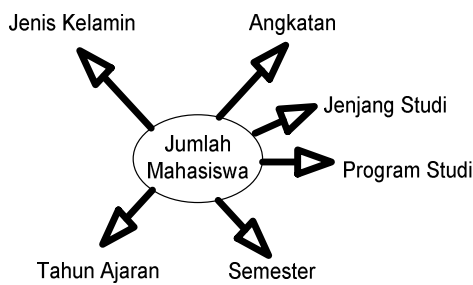

Gambar 4. *Hypercube* laporan jumlah mahasiswa aktif per semester tahun ajaran per jenjang program studi per jenis kelamin per angkatan

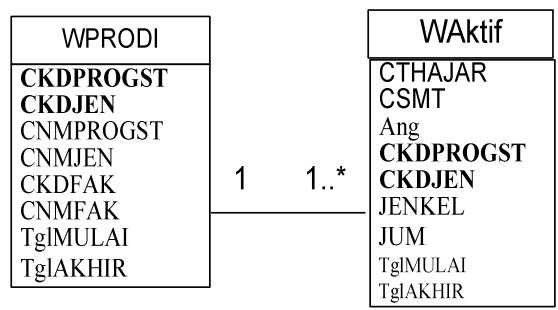

Gambar 5. *Class diagram data warehouse* laporan jumlah mahasiswa aktif per semester tahun ajaran per jenjang program studi per jenis kelamin per angkatan

Gambar 4 memperlihatkan hpercube laporan jumlah mahasiswa aktif per semester tahun ajaran per jenjang program studi per jenis kelamin per angkatan. Berdasarkan *hypercube* Gambar 4 maka akan terbentuk 1 tabel fakta dan 6 tabel dimensi, dan sesuai dengan penjelasan sebelumnya dimensi Jenis kelamin, angkatan, jenjang studi tidak perlu dibuat.. Untuk dimensi tahun ajaran juga tidak perlu membuat tabel dimensi oleh karena hanya berisi data tahun ajaran, demikian juga untuk dimensi Semester tidak perlu membuat tabel dimensi oleh karena hanya berisi dengan 2 nilai yaitu semester Ganjil/Gasal atau semester Genap. Oleh karena itu akan terbentuk 1 tabel fakta dan 1 tabel dimensi, hal ini akan terlihat pada model data Gambar 5. Tabel fakta adalah WAktif dan tabel dimensi adalah WPRODI. Tabel dimensi WPRODI adalah penggabungan 3 tabel pada sistem *OLTP* aitu tabel MPRODI, MFAKULTAS dan MJENJANG. Jika menggunakan d*ata warehouse* maka laporan ini membutuhkan 2 tabel basis data yaitu WAktif dan WPRODI dimana jika menggunakan database OLTP menggunakan 5 tabel yaitu MMAHASISWA, TRKRS, MPRODI, MFAKULTAS dan MJENJANG.

Gambar 6 memperlihatkan *hypercube* laporan jumlah komposisi indeks prestasi per semester tahun ajaran per jenjang program studi per angkatan. Berdasarkan *hypercube* Gambar 6 maka akan terbentuk 1 tabel fakta dan 6 tabel dimensi, dan sesuai dengan penjelasan sebelumnya dimensi angkatan, tahun ajaran, jenjang studi dan semester tidak perlu dibuat. Untuk dimensi Kurang/Cukup/Baik hanya akan berisi dengan 3 nilai yaitu K untuk data yang mempunyai IPS (Indeks Prestasi Semester) kurang dari 2.5, C untuk data yang mempunyai IPS (Indeks Prestasi Semester) antara 2.5 sampai 3.0, B untuk data yang mempunyai IPS (Indeks Prestasi Semester) lebih dari 3.0. Oleh karena itu tidak perlu membuat tabel dimensi untuk dimensi Kurang/Cukup/Baik ini. Oleh karena itu akan terbentuk 1 tabel fakta dan 1 tabel dimensi, hal ini akan terlihat pada model data Gambar 7. Tabel fakta adalah WIPS dan tabel dimensi adalah WPRODI. Tabel dimensi WPRODI adalah penggabungan 3 tabel pada sistem *OLTP* yaitu tabel MPRODI, MFAKULTAS dan MJENJANG. Jika menggunakan d*ata warehouse* maka laporan ini membutuhkan 2 tabel basis data yaitu WIPS dan WPRODI dimana jika menggunakan database OLTP menggunakan 4 tabel yaitu TRKRS, MPRODI, MFAKULTAS dan MJENJANG.

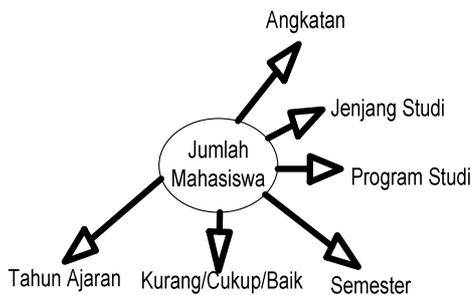

Gambar 6. *Hypercube* laporan jumlah komposisi indeks prestasi per semester tahun ajaran per jenjang program studi per angkatan

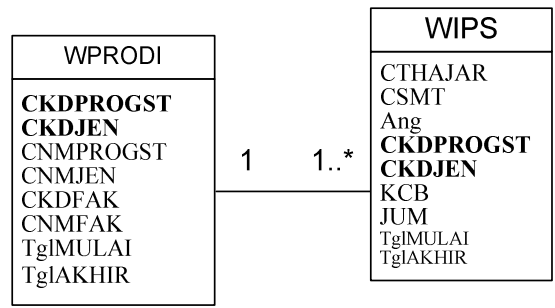

Gambar 7. *Class diagram data warehouse* laporan jumlah komposisi indeks prestasi per semester tahun ajaran per jenjang program studi per angkatan

Gambar 8 memperlihatkan hypercube laporan jumlah komposisi grade nilai per semester tahun ajaran per jenjang program studi per jenis kelamin per angkatan. Berdasarkan *hypercube* Gambar 8 maka akan terbentuk 1 tabel fakta dan 7 tabel dimensi, dan sesuai dengan penjelasan sebelumnya dimensi jenis kelamin, angkatan, tahun ajaran, jenjang studi dan semester tidak perlu dibuat. Untuk dimensi Grade hanya akan berisi dengan 6 nilai yaitu Grade A, B, C,D, E, dan – untuk yang tidak mengikuti ujian, maka tidak perlu membuat tabel dimensi untuk dimensi Grade ini. Oleh karena itu akan terbentuk 1 tabel fakta dan 1 tabel dimensi, hal ini akan terlihat pada model data Gambar 9. Tabel fakta adalah WGrade dan tabel dimensi adalah WPRODI. Tabel dimensi WPRODI adalah penggabungan 3 tabel pada sistem *OLTP* yaitu tabel MPRODI, MFAKULTAS dan MJENJANG. Jika menggunakan d*ata warehouse* maka laporan ini membutuhkan 2 tabel basis data yaitu WGrade dan WPRODI dimana jika menggunakan database OLTP menggunakan 5 tabel yaitu MMAHASISWA, TRKRS, MPRODI, MFAKULTAS dan MJENJANG.

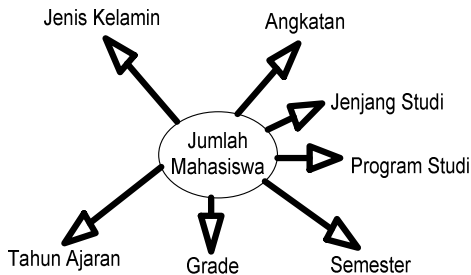

Gambar 8. *Hypercube* laporan jumlah komposisi grade nilai per semester tahun ajaran per jenjang program studi per jenis kelamin per angkatan

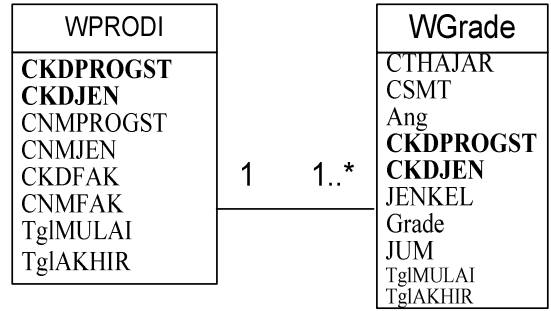

Gambar 9. *Class diagram data warehouse* laporan jumlah komposisi grade nilai per semester tahun ajaran per jenjang program studi per jenis kelamin per angkatan

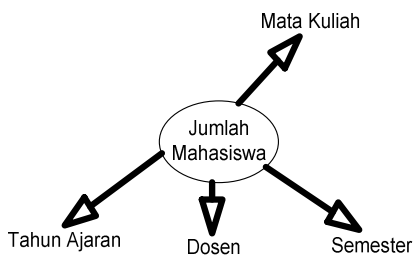

Gambar 10. *Hypercube* laporan pengajaran dosen per semester tahun ajaran

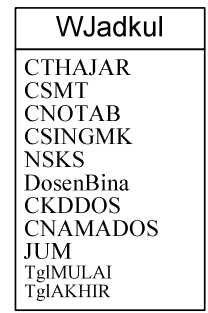

Gambar 11. *Class diagram data warehouse* laporan pengajaran dosen per semester tahun ajaran

Gambar 10 memperlihatkan hypercube laporan pengajaran dosen per semester tahun ajaran. Berdasarkan *hypercube* Gambar 10 maka akan terbentuk 1 tabel fakta dan 4 tabel dimensi, dan sesuai dengan penjelasan sebelumnya dimensi tahun ajaran dan semester tidak perlu dibuat. Untuk dimensi dosen tidak perlu dibuat tabel dimensi oleh karena laporan ini hanya menampilkan nama dosen,maka nama dosen tersebut disertakan pada tabel fakta. Demikian juga dengan dimensi mata kuliah tidak perlu dibuat tabel dimensi oleh karena laporan ini hanya menampilkan nama mata kuliah dan SKS (Satuan Kredit Semester) mata kuliah serta nama dosen pembina, maka nama singkatan mata kuliah dan SKS (Satuan Kredit Semester) mata kuliah serta nama dosen Pembina disertakan pada tabel fakta. Oleh karena itu akan terbentuk 1 tabel fakta dan tidak ada tabel dimensi, hal ini akan terlihat pada model data Gambar 11. Jika menggunakan d*ata warehouse* maka laporan ini membutuhkan hanya 1 tabel basis data yaitu WJadkul dimana jika menggunakan database OLTP menggunakan 4 tabel yaitu TJADKUL, TDOSFAK, MTBMTK dan MFAKULTAS.

Akhirnya pada masing-masing tabel di atas, baik tabel fakta atau dimensi ditambahkan *field* tglmula dan tglakhir yang berfungsi sebagai proses pembaharuan data *record*, oleh karena pada d*ata warehouse* tidak boleh dilakukan proses penghapusan record. Data record yang masih berlaku adalah apabila *field* tglakhir masih kosong, apabila *field* tglakhir telah terisi maka akan terbentuk *record* duplikat yang menggambarkan data *record* yang terkini.

## 2.3. Model data logika data warehouse

Langkah ketiga pada metode penelitian ini adalah mendesain model data logika data warehouse berdasarkan hypercubes bentukan di atas dan secara keseluruhan model data logika d*ata warehouse* yang terbentuk berdasar analisa laporan-laporan di atas yang menggunakan konsep dimensi bisnis dengan pendekatan kubus multidimensi *hypercube terlihat* pada Gambar 12.

Langkah kelima pada metode penelitian ini adalah pengimplementasian database data warehouse dengan menggunakan aplikasi database MySQL dan dimana sebelumnya langkah keempat telah dilakukan dimana database OLTP diimplementasikan juga dengan aplikasi database MySQL. Langkah keenam pada metode penelitian ini dilakukan dengan menjalankan query untuk menghasilkan record yang akan menampilkan laporan dengan menggunakan database OLTP dan demikian pula dengan langkah ketujuh dimana dilakukan dengan menjalankan query untuk menghasilkan record yang akan menampilkan laporan dengan menggunakan data warehouse.

## 3. HASIL DAN PEMBAHASAN

Untuk membuktikan bahwa penggunaan data warehouse lebih efektif daripada database pada umumnya, maka akan dibuktikan dari efisiensi pada total byte yang dikelola, record yang dikelola, panjang record yang diproses, jumlah tabel yang diproses, waktu dan record yang dihasilkan. Tabel 1 merupakan besaran 8 tabel database OLTP yang terdapat pada gambar 1, yang mempunyai total panjang record 1,099 byte, total jumlah record 131,171 record dan total keseluruhan byte adalah 31,303,511 byte. Sedangkan Tabel 2 merupakan besaran 6 tabel data warehouse yang terdapat pada Gambar 12, yang mempunyai total panjang record 326 byte, total jumlah record 1,138 record dan total keseluruhan byte adalah 71,555 byte.

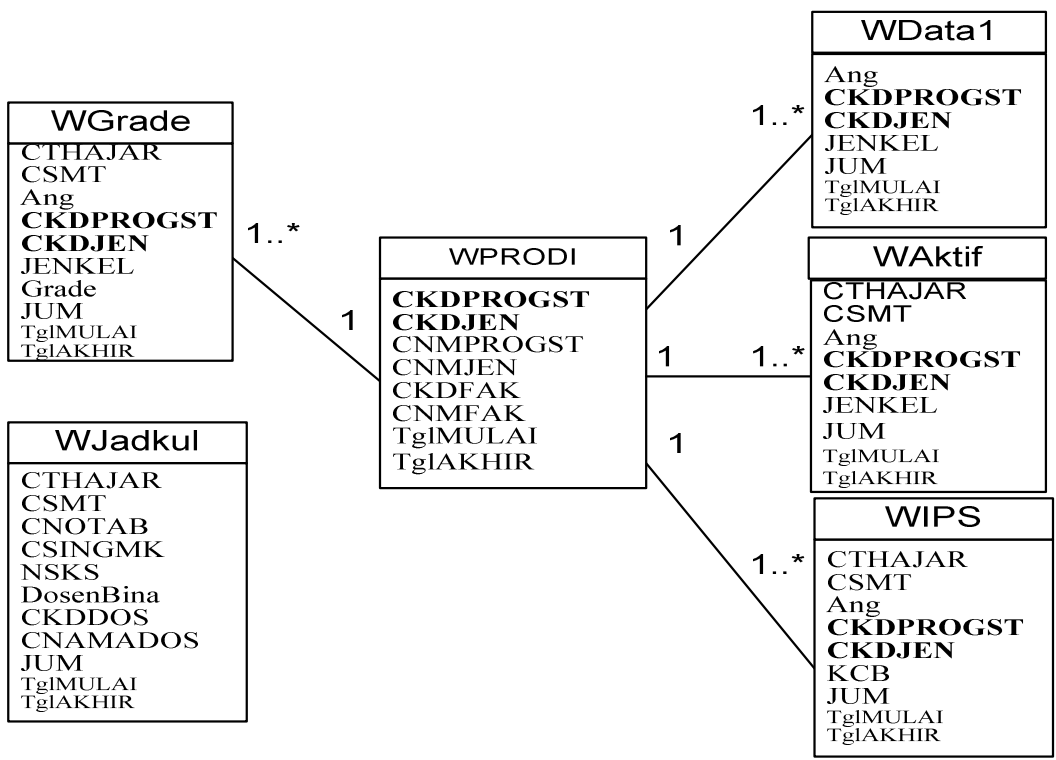

Gambar 12. *Class diagram* model data logika d*ata warehouse*

Tabel 3 memperlihatkan prosentase efisiensi dimana penggunaan data warehouse lebih efisien 237.12% (1099-326)/326*100 untuk panjang record, lebih efisien 11,426.45% (131,171-1138)/1138*100 untuk jumlah record dan lebih efisien 43,467.48% (31,303,511-71,555)/71,555*100 untuk total keseluruhan byte. Prosentase efisiensi ini diukur sesuai dengan langkah kesembilan pada metode penelitian ini dimana nilai kuantitatif ini diukur dengan prosentase kenaikan dengan rumus : (data lama – data baru)/ data baru *100.

Tabel 1. Tabel besaran isi tabel database transaksional

| Nama Tabel | Panjang record | Jumlah record | Total byte |
|---|---|---|---|
| MMAHASIWA | 586 byte | 42977 record | 25 184 522 byte |
| MPRODI | 48 byte | 16 record | 768 byte |
| MFAKULTAS | 65 byte | 7 record | 455 byte |
| MJENJANG | 24 byte | 3 record | 72 byte |
| TRKRS | 68 byte | 84774 record | 5 764 632 byte |
| TJADKUL | 88 byte | 1988 record | 174 944 byte |
| TDOSFAK | 73 byte | 386 record | 28 178 byte |
| MTBMTKL | 147 byte | 1020 record | 149 940 byte |
| Total | 1099 byte | 131171 record | 31303511 byte |

Tabel 2. Tabel besaran isi tabel data warehouse

| Nama Tabel | Panjang record | Jumlah record | Total byte |
|---|---|---|---|
| WPRODI | 35 byte | 16 record | 560 byte |
| WJADKUL | 127 byte | 303 record | 38481 byte |
| WGRADE | 44 byte | 368 record | 16192 byte |
| WDATA1 | 34 byte | 279 record | 9486 byte |
| WAKTIF | 43 byte | 74 record | 3182 byte |
| WIPS | 43 byte | 98 record | 4214 byte |
| Total | 326 byte | 1138 record | 71555 byte |

Tabel 3. Tabel total perbandingan kapasitas database transaksional dan data warehouse

| Variabel | Panjang record | Jumlah record | Total byte |
|---|---|---|---|
| Total database OLTP | 1099 byte | 131171 record | 31303511 byte |
| Total data warehouse | 326 byte | 1138 record | 71555 byte |
| Prosentase efisiensi perbandingan database/DW | 237.12 % | 11.426,45% | 43.647,48% |

Sesuai dengan langkah kedelapan pada metode penelitian ini dimana Tabel 4 memperlihatkan bahwa penggunaan data warehouse lebih efisien pada total byte yang dikelola, record yang dikelola, panjang record yang diproses, jumlah tabel yang diproses, waktu dan record yang dihasilkan pada kelima laporan di atas. Prosentase efisiensi pada Tabel 4 ini diukur sesuai dengan langkah kesembilan pada metode penelitian ini dimana nilai kuantitatif ini diukur dengan prosentase kenaikan dengan rumus: (data lama – data baru)/data baru*100. Berikut ini penjelasan dasar pembentukan masing-masing pengukuran prosentase efisiensi:
1) Total byte yang dikelola didapatkan dari total perkalian record yang dikelola dan panjang record yang dikelola dari tabel database yang digunakan dalam menghasilkan sebuah laporan baik dengan menggunakan database OLTP atau data warehouse.
2) Record yang dikelola didapatkan dari total jumlah record yang diproses dalam menghasilkan laporan baik dengan menggunakan database OLTP atau data warehouse.
3) Panjang record yang dikelola didapatkan dari total panjang byte record yang diproses dalam menghasilkan laporan baik menggunakan database OLTP atau data warehouse.
4) Jumlah Tabel yang digunakan didapatkan dari penjelasan bab sebelumnya di atas dalam menghasilkan laporan baik dengan menggunakan database OLTP atau data warehouse.
5) Waktu proses didapatkan dari pengujian pengeksekusian query dalam menghasilkan record untuk membentuk masing-masing laporan dengan menggunakan aplikasi database MySql, baik dengan menggunakan database OLTP ataupun dengan data warehouse.
6) Record yang dihasilkan didapatkan dari hasil eksekusi query pada masing-masing laporan untuk membentuk laporan tersebut dengan menggunakan aplikasi database MySQL, baik dengan menggunakan database OLTP ataupun dengan data warehouse, dimana hasil record tersebut mempunyai hasil yang sama pada beberapa laporan.

Sebagai contoh laporan 1 adalah laporan jumlah mahasiswa per jenjang program studi per jenis kelamin per angkatan. Dimana pada pada Tabel 4 dibawah laporan ini jika dihasilkan dengan menggunakan data warehouse akan mempunyai prosentase efisiensi sebagai berikut :
1) Prosentasi efisiensi total byte yang dikelola sebanyak 250,605.31% (25,185,855-10,046)/10.046*100.
2) Prosentase efisiensi record yang dikelola sebanyak 14,477.63% (43,004-295)/295*100.
3) Prosentase efisiensi panjang record yang diproses sebanyak 947.83% (723-69)/69*100.
4) Prosentase efisiensi jumlah tabel yang diproses sebanyak 100% (4-2)/2*100.
5) Prosentasi efisiensi waktu per detik sebanyak 31,200% (3.13-0.01)/0.01*100.
6) Prosentase efisiensi jumlah record yang dihasilkan sebanyak 0% (279-279)/279*100.

Berdasarkan Tabel 4, semua efisiensi mengalami kenaikan prosentase, namun untuk efisiensi *record* yang dihasilkan pada laporan 1, 2 dan 3 tidak mengalami kenaikan prosentase. Prosentase kenaikan efisiensi tertinggi adalah 5,008,200% pada efisiensi waktu pada laporan 4 dan kenaikan efisiensi terendah adalah 95.38%, yaitu pada efisiensi *record* yang dihasilkan pada laporan 5.

Tabel 4. Tabel efisiensi perbandingan penggunaan database dan data warehouse

|  |  | Laporan 1 | Laporan 2 | Laporan 3 | Laporan 4 | Laporan 5 |
|---|---|---|---|---|---|---|
| Total Byte | Database | 25.185.855,00 | 30.255.131,00 | 5.040.609,00 | 30.225.131,00 | 352.023,00 |
|  | Datawarehouse | 10.046,00 | 3.742,00 | 4.774,00 | 16.752,00 | 38.481,00 |
|  | Efisiensi → | 250.605,31% | 808.428,35% | 105.484,60% | 180.327,00% | 814,80% |
| Record yang Dikelola | Database | 43.004,00 | 117.111,00 | 74.134,00 | 117.111,00 | 3.384,00 |
|  | Datawarehouse | 295,00 | 90,00 | 114,00 | 384,00 | 303,00 |
|  | Efisiensi → | 14.477,63% | 130.023,33% | 64.929,82% | 30.397,66% | 1.016,83% |
| Panjang Record | Database | 723,00 | 791,00 | 205,00 | 791,00 | 373,00 |
|  | Datawarehouse | 69,00 | 78,00 | 78,00 | 79,00 | 127,00 |
|  | Efisiensi → | 947,83% | 914,10% | 162,82% | 901,27% | 193,70% |
| Tabel yang digunakan | Database | 4 | 5 | 4 | 5 | 4 |
|  | Datawarehouse | 2 | 2 | 2 | 2 | 1 |
|  | Efisiensi → | 100,00% | 150,00% | 100,00% | 150,00% | 300,00% |
| Waktu/Detik | Database | 3,13 | 245,61 | 439,68 | 500,83 | 119,15 |
|  | Datawarehouse | 0,01 | 0,01 | 0,01 | 0,01 | 0,01 |
|  | Efisiensi → | 31.200,00% | 2.456.000,00% | 4.396.700,00% | 5.008.200,00% | 1.191.400,00% |
| Record yang dihasilkan | Database | 279,00 | 74,00 | 3091,00 | 368,00 | 592,00 |
|  | Datawarehouse | 279,00 | 74,00 | 98,00 | 368,00 | 303,00 |
|  | Efisiensi → | 0,00% | 0,00% | 3.054,08% | 0,00% | 95,38% |

## 4. SIMPULAN

Data warehouse mempengaruhi proses transaksi pada database dimana database data warehouse yang dihasilkan akan lebih ramping dan perintah sql yang dijalankan untuk mengakses data warehouse secara nyata akan lebih cepat, yaitu jumlah record yang diproses makin sedikit dan proses join berkurang. Dari keseluruhan efisiensi kenaikan prosentase jika digabungkan akan menghasilkan rata-rata efisiensi kenaikan prosentase 461.801,84%. yang menunjukkan bahwa penggunaan data warehouse lebih handal dan efisien dibandingkan penggunaan database OLTP.

## DAFTAR PUSTAKA

[1]. Silva FSC. Panigassi R. Hulot C. Learning Management Systems Desiderata for Competitive Universities. *European Journal of Open Distance and E-Learning*. 2007; 13(2): 121-129.
[2]. Ward J. Peppard J. *Strategic planning for Information Systems*. Third Edition. West Susse.: John Willey & Sons Ltd. 2003.
[3]. Porter ME. Strategy and the Internet. *Harvard Business Review*. 2001; 79(3): 62-78.
[4]. Dimokas N. Mittas N. Nanopoulos A. Angelis L. *A Prototype System for Educational Data Warehousing and Mining*. Proceedings of the 2008 Panhellenic Conference on Informatics. 2008: 199-203.

[5]. Wikramanayake GN. Goonetillake JS. Managing Very Large Databases and Data Warehousing. *Sri Lankan Journal of Librarianship and Information Management*. 2006;2(1):22-29.
[6]. Goldstein PJ. Karzt RN. Academic Analytics: The Uses of Management Information and Technology in Higher Education. *EDUCAUSE Review*. 2005; 7(1): 1-12.
[7]. Hans D. Gomez JM. Peters D. Solsbach A. Case Study-Design for Higher Education-A Demonstration in the Data Warehouse Environment. in: Abramowicz W. Flejter D (Editors). BIS 2009 Workshop. *LNBIP*. 2009; 37: 231-241.
[8]. Wu T. System of Teaching Quality Analyzing and Evaluating Based on Data Warehouse. Computer Engineering and Design. 2009; 30(6): 1545-1547.
[9]. Zhou L. Wu M. Li S. *Design of Data Warehouse in Teaching State Based on OLAP and Data Mining*. Proc. SPIE (The International Society for Optical Engineering). 2009; 7344: 23-29.
[10]. Gombiro C. Munyoka W. Hove S. Chengetanai G. Zano C. The Need for Data Warehousing in Sharing Learning Materials. *Journal of Sustainable Development in Africa*. 2008; 10(2): 422-449.
[11]. Ranjan J. Khalil S. Conceptual Framework of Data Mining Process in Management Education in India: An Institutional Perspective. *Information Technology Journal*. 2008; 7(1): 16-23.
[12]. Calero C. Piatiini M. Pascual C. Serrano MA. *Towards Data Warehouse Quality Metrics*. Proceedings of the 3rd Intl. Workshop on Design and Management of Data Warehouses (DMDW'2001). Interlaken. Switzerland. 2001; 39: 2-11.